
\magnification=1200
\baselineskip=20pt
\def\lsim{<\kern-2.5ex\lower0.85ex\hbox{$\sim$}\ }
\def\rsim{>\kern-2.5ex\lower0.85ex\hbox{$\sim$}\ }

\overfullrule=0pt
\centerline{\bf  NEUTRON ACCELERATION IN UNIFORM
ELECTROMAGNETIC FIELDS}
\vskip .5cm
\centerline{by}
\vskip .5cm
\centerline{J. Anandan}
\centerline{Department of Physics and Astronomy}
\centerline{University of South Carolina}
\centerline{Columbia, SC 29208}
\vskip .5cm
\centerline{and}
\vskip .5cm
\centerline{C. R. Hagen}
\centerline{Department of Physics and Astronomy}
\centerline{University of Rochester}
\centerline{Rochester, NY 14627}
\vskip 1cm
\centerline{\bf  \underbar{Abstract}}
The question as to whether neutron acceleration can occur in uniform
 electromagnetic fields is examined.  Although such an effect has
been predicted using the canonical equations of motion some doubt
has been raised recently as to whether it is in principle observable
for a spin 1/2 particle.  To resolve this issue a gedanken experiment
is proposed and analyzed using a wave packet construction for the
neutron beam.  By allowing arbitrary orientation for the neutron spin
as well as for the electric and magnetic fields a non vanishing
acceleration of the center of the neutron wave packet is found which
confirms the predictions of the canonical formalism.

\vfil\eject
It is well known that a particle with a magnetic dipole moment such
as a neutron must undergo an acceleration in an inhomogeneous
magnetic field. More recently it has been
 shown$^{1,2}$ that a magnetic
dipole
in homogeneous magnetic and electric fields must undergo an
acceleration both classically and quantum mechanically. This a
consequence of the Hamiltonian$^{3,4}$
$$H = {1 \over 2m} {\bf p}^2 - {\bf \mu} \cdot [ {\bf  B} - {1 \over
mc}\
{\bf  p} \times {\bf  E} ]\eqno(1)$$
where ${\bf  p}$ and $m$ refer to the momentum and mass
respectively of the
particle (e.g., a neutron).  The magnetic moment is ${\bf  \mu} =
\gamma
{\bf  \sigma}/2$ with ${\bf  \sigma}$
 being the set of Pauli spin matrices.  It is
assumed that the fields ${\bf  E}$ and ${\bf  B}$ are uniform and time
independent.  The Hamiltonian (1) has been obtained as a low energy
consequence of Lorentz covariance$^2$ and may therefore be
regarded as generally valid in  both the classical and quantum mechanical
cases
whenever the magnetic moment of a neutral particle arises from its
spin angular momentum.
Upon calculating the commutators (or Poisson brackets) of $H$ with ${\bf  p}$,
 ${\bf  x}$, and ${\bf  \sigma}$ one finds
$$\eqalignno{ {\bf  \dot x} &= {\bf  p}/m + {1 \over mc}\ {\bf  E}
\times
{\bf  \mu} &(2)\cr
{\bf  \dot p} &= 0 &(3)\cr
\dot{{\bf  \sigma}} &= \gamma {\bf  \sigma} \times [{\bf  B} - {1
\over mc}\
{\bf  p} \times {\bf  E}] \quad  .&(4)\cr}$$
It follows upon insertion of (4) into the time derivative of (2)
that$^1$
$$\ddot{{\bf  x}} = {\gamma \over mc}\ {\bf  E} \times ({\bf  \mu}
\times
{\bf  B}) + O(E^2) \quad  .\eqno(5)$$
Thus there is a nonvanishing
acceleration proportional in lowest order to both $\vert {\bf  E}
\vert$ and
$\vert {\bf  B} \vert$. For $E=10^7 V/m$ and $B=1T$ this can be as big as $6$
 cm/sec$^2$. The difference between the canonical
momentum ${\bf  p}$
and the kinetic momemtum $m {\bf  v}$ is sometimes called the \lq\lq hidden
momentum", which has been studied  also by others$^5$.
The acceleration (5) arises from the variation of the
hidden momentum which follows from the precession of the magnetic moment in
the magnetic field.

Although the overall consistency of the canonical formalism of
quantum
mechanics would appear to offer no alternative to the result (5),
there has
been a recent suggestion$^6$ that in fact the predicted acceleration is
unobservable even in principle.  It should be noted, however, that at
least under the assumption of constant fields such a question must be
capable of being
resolved unambiguously by a direct calculation based on
Schr\"odinger's
equation.  In particular a neutron which passes from a field free
region to
one described by the Hamiltonian (1) can be viewed as being subject
to a
constant but spin dependent potential.  The problem then reduces to
solving Schr\"odinger's equation for motion through this potential for
fixed ${\bf  p}$.

More specifically, one may imagine carrying out the following
experiment.
 A wave packet describing a neutron is allowed to propagate in the
field
free region $z < 0$ and to enter the uniform field region $z >0$ at
normal
incidence.  The coordinate system is chosen so that the center of the
wave
packet passes through the origin at time $t=0$. The neutron is
assumed to be totally polarized in the direction of the unit vector
${\bf  n}$.  At a distance $z_D$ from the origin a detector is placed
which
measures the transverse displacement of the beam as a function of
$z_D$ to
arbitrary accuracy.  Since there can be no possibility of carrying out
such
measurements without a transverse localization of the beam, it is
evident
that the wave packet must be spread in at least one of the two
transverse momentum variables which we take without loss of
generality to be $p_x$.  Furthermore, a spreading in the $z$
coordinate is required in order to allow a time of flight to be
 inferred from the detector position $z_D$.

Thus the spatial part of the wave function for $z < 0$, $t < 0$ is given
by
$$\eqalign{\psi (x,z) = &\int {dp_z dp_x \over
(2 \pi)^2}\,\exp(ip_xx+ i p_z z)\cr
&\exp \left( -i { p_x^2 + p_z^2 \over
2m}t\right) f(p_x,p_z)\quad .\cr}\eqno(6)$$
The momentum space wave function $f(p_x, p_z )$ is taken to be an even function
 in $p_x$ which is peaked around the point
$p_x = 0$, $p_z = k$.  It is normalized by the condition
$$\int {dp_z dp_x \over (2 \pi )^2}\
\vert f (p_x,p_z) \vert^2 = 1 \quad  .$$
While a Gaussian function $$f (p_x, p_z)=\left({2\pi\over \Delta p_x \Delta
 p_z}\right)^{1\over2}exp\left(-{p_x^2\over {4\Delta p_x}^2}-{(p_z-k)^2\over
 {4\Delta p_z}^2}\right)$$ would allow an
explicit calculation of the wave function to be performed, it is not in
fact required for this problem.  A localization
of the wave function in the y-
direction is possible as well, but is basically irrelevant to the result.

When the wave packet passes through the origin the usual reflection
and
transmission effects are encountered although one clearly is
interested
only in the transmitted part for the purpose of this work.  Matching
the
function and its derivative at $z = 0$ one finds that
 to lowest order in the fields the spin part is
unaltered while the transmitted beam has a form identical to that
given by
(6) provided that $p_z$ in the exponential $e^{ip_z z}$ is replaced by
the
momentum component $\tilde p_z$ appropriate to propagation in the
nonzero
field region.  The latter is obtained from the equation
$${p^2_z \over 2m} = {\tilde p_z^2 \over 2m} - {\bf \mu} \cdot \left[
{\bf  B} -
{1 \over mc}\ {\bf  p} \times {\bf  E} \right] \eqno(7)$$
where (to lowest nonvanishing order in ${\bf  E}$) it is the vector
${\bf  p}$
which appears on the right hand side of (7) in combination with
${\bf  E}$.
Clearly, one could proceed at this point by considering separately the
two
eigenmodes of propagation and determining the appropriate spin
part of the
wave function by defining a spin basis with respect to the direction
of the magnetic field in the \lq\lq neutron rest frame"
$${\bf  B'} = {\bf  B} - {1 \over mc}{\bf  p} \times {\bf  E}.$$   Fortunately,
 a simpler
and more elegant approach is possible which involves calculating
$\tilde p_z$ as a matrix in the spin space and using the projector or
the density matrix in spin space
$$P_n = {1 \over 2}\ \left( 1 + {\bf \sigma} \cdot {\bf  n} \right),$$
satisfying ${P_n}^2 = P_n$, to include the initial polarization state of the
 beam.

To the required order one finds from (7) that
$$\tilde p_z = p_z + {m \gamma \over 2 p_z}\ {\bf  \sigma} \cdot
\left[ {\bf  B} - {1 \over mc}\ {\bf  p} \times {\bf  E} \right] \quad  .$$
This then leads to the evaluation of
$$\eqalign{<x>
&= \int dx dz \quad x
\int {dp_z dp_z^\prime  dp_xdp_x^\prime
 \over (2 \pi)^4}
\quad  e^{i( p_x - p_x^\prime) x + i(p_z - p_z^\prime)z}\cr
&\quad  \exp \bigg( -i\ {p_x^2 +
 p^2_z - {p^\prime_x} ^2 -p^{\prime 2}_z \over 2m} \ t \bigg)\quad
  f (p_x,p_z )
 f^* ( p_x^\prime,p^\prime_z)
 T\cr}\eqno(8)$$ where
$$\eqalign{T = \ {\rm Tr}\ &\exp \bigg[ i z \ {m \gamma \over 2p_z}
{\bf  \sigma} \cdot
\big( {\bf  B} - {1 \over mc}
 \ {\bf  p} \times {\bf  E}) \bigg] \ {1 \over 2}\ (1 +
{\bf  \sigma} \cdot {\bf  n})\cr
&\exp \bigg[ - i z \ {m \gamma \over 2 p^\prime_z}\
{\bf  \sigma} \cdot \big( {\bf  B}
 - {1 \over mc}\ {\bf  p}^\prime \times {\bf  E} \big) \bigg] \quad .\cr}
\eqno(9)$$
The evaluation of the trace is simplified by the observation that
because
of the antisymmetry of the integrand in Eq. (8) under the
simultaneous
change of sign of $ x$, $p_x$, and
$p_x^\prime$ only those terms in $T$ which are odd in
 $p_x$ and $p_x^\prime$ can contribute.  One proceeds by writing
$$\eqalign{\exp \bigg[ i z \ {m \gamma \over 2p_z}\ {\bf  \sigma}
\cdot
\big( {\bf  B} - &{1 \over mc}\ {\bf  p} \times {\bf  E} \big) \bigg]\cr
&= C + i \vert {\bf  B} - {1 \over mc}\ {\bf  p} \times {\bf  E}
\vert^{-1} {\bf  \sigma} \cdot \big( {\bf  B}
 - {1 \over mc}\ {\bf  p} \times {\bf  E}
\big)S\cr}$$
where
$$(C,S) \equiv (\cos, \sin) \bigg[ {m \gamma z \over 2 p_z}\
\vert {\bf  B} - {1 \over mc}\ {\bf  p} \times {\bf  E} \vert \bigg]$$
which is easily seen by a diagonalization of the matrix in the exponent.
Using a prime to denote the same quantities when $p$ is
replaced by
$p^\prime$, it is first noted that the $CC^\prime$
 term makes no contribution.
To the desired order one thus finds for (9) the result
$$\eqalign{T &= {i \over \vert {\bf  B} \vert^2} \ SS^\prime {\bf  n}
\times
{\bf  B} \cdot {\bf  E} \times ({\bf  p} - {\bf  p}^\prime )\ {1 \over
mc}\cr
&\quad  + {i \over 2}\ (CS^\prime + C^\prime S) \ {1 \over
\vert {\bf  B} \vert}\ {\bf  n} \cdot
{\bf  E} \times ({\bf  p} - {\bf  p}^\prime ) \ {1 \over mc}\quad  .\cr}$$

Upon inserting this into (8) it is observed that the canonical
commutation
relations imply that the combination $x({\bf  p}
 - {\bf  p}^\prime)$ becomes $i {\bf \ell}$, where ${\bf \ell}$ is the unit
 vector in the x-direction. Hence, (8) may be written as
$$\eqalign{< x >\ &= \int {dp_x dp_z \over
(2 \pi)^2}\ \vert f(p_x, p_z) \vert^2 \cr
& \quad  \bigg[ - {1 \over mc \vert {\bf  B} \vert^2}\ S^2 ({\bf  n}
 \times {\bf  B}) \cdot
({\bf  E} \times {\bf \ell}) - {1 \over mc \vert {\bf  B} \vert}\
CS {\bf  n} \cdot {\bf  E} \times {\bf \ell} \bigg] \quad .\cr}\eqno(10)$$
To complete the calculation one notes that it is sufficient to work to
lowest order in ${\bf  E}$ and ${\bf  B}$ and to neglect corrections of
the order
of $\Delta p_z/k$ where $\Delta p_z$ is the wave packet width in
momentum
space.  This allows $C$ and $S$ to be replaced by 1 and $m \gamma z
\vert
{\bf  B} \vert /2k$, respectively.  One now regards $\bf {\ell}$, which
is in the transverse direction in which the wave is localized, to be an
arbitrary unit vector in the xy-plane so that $x$ is replaced by
${\bf x} \cdot {\bf {\ell}}$. Then (10) reads
$$\eqalign{{< \bf x} \cdot {\bf {\ell}}> &= \big\{ {1 \over 2}\ t^2 \ {\gamma
\over
mc}\ {\bf {\ell}} \cdot \big[ {\bf  E} \times (\gamma \ {1 \over
2}\
{\bf  n} \times {\bf  B}) \big] \cr
&\quad  + t {\bf {\ell}} \cdot {\bf  E} \times {1 \over 2} \ \gamma
{\bf  n} \
{1 \over mc}\big\}\cr}\eqno(11)$$
where $z$ has been replaced by $z_D$ which in turn is related to the
time
of flight $t$ by $z_D = kt/m$.  This identification is made possible by
 the fact that the wave packet is localized in the $z$ coordinate
while the velocity of its center may be taken in eq. (11) to be $k/m$
in the present approximation. Eq. (11) is valid for small $t$ meaning
that $t<<$(1/Larmor frequency). This restriction is needed because the
acceleration averages to zero for large $t$, which was why it was
proposed$^2$ to observe it by sending the neutrons through an array
with periodic field reversals which would allow
the effect of the acceleration to
accumulate.

The verification of the expressions derived in ref. 1 for the
acceleration
and the kinetic momentum is now immediate.  For short times $t$ Eq.
(5)
clearly implies that there should exist a $t^2$ term in the mean
transverse
displacement whose coefficient is one-half the acceleration.  It is
striking that the calculation presented here yields an acceleration
which
has precisely the vector structure implied by the canonical
formalism.
Also noteworthy is the fact that the second term in (11) is linear in
$t$
and corresponds to the uniform drift of the particle beam implied by
the
difference between the canonical and kinetic momenta as indicated
in Eq. (2).
This completes the rederivation in the Schr\"odinger picture of the
velocity and acceleration obtained previously in the Heisenberg
picture$^1$ and classically$^{1,2}$.

The recent
claim$^6$ that this acceleration is not observable for a spin 1/2
particle will now be examined.
 This analysis, as will be seen, illustrates the
physical meaning of the hidden momentum and shows interesting
aspects of the velocity dependent potential in (1). This claim is
based on the plane wave solution obtained in ref. 4 which may be
written as
$$\psi = A exp(ik_+z)|+\rangle + B exp(ik_-z)|-\rangle ,\eqno(12)$$
where $|+\rangle$ and $|-\rangle$ are respectively the spin states
with respect to a quantization axis along ${\bf  B'}$.
It was claimed$^6$ that for each of these two states, the acceleration
(5) vanishes within the accuracy of the Hamiltonian (1). Indeed,
in the computation of the phase shift experienced by the neutron in
ref. 4 two separate phase shifts were obtained for these two
eigenmodes, and
this prediction was experimentally confirmed$^7$. The correctness of
this analysis is ultimately due to the following fact.
Suppose $F_C$ is the operator
 which depends on external fields, mirrors, and the
space-time path $C$ around which the interference takes
place
that transforms one of the interfering
states in the interfering region into the other.
 The observation of the
variation of intensity in an interference experiment for different
incoming states determines the eigenvalues of the Hermitian
operator$^8$ $R =F_C+F_C^\dagger$.
In the present case,
$|+\rangle$ and $|-\rangle$ are the eigenstates of this operator R.

However, to study the acceleration experienced by the neutron,
it is necessary to express $\psi$ in terms of the {\it eigenstates of the
quantum mechanical acceleration operator} defined by (5). For the
neutron, this acceleration operator is
$$\ddot{{\bf  x}} = {\gamma\mu \over mc}\ {\bf  E} \times ({\bf
\sigma} \times{\bf  B}) $$
whose eigenstates are denoted by $|\uparrow\rangle$ and
$|\downarrow\rangle$. The corresponding eigenvalues are
$\pm|{\gamma\mu \over mc}\ {\bf  E} \times ({\bf  \sigma}
\times{\bf  B})|$, which are therefore the accelerations experienced
by the eigenmodes that are proportional to $|\uparrow\rangle$ and
$|\downarrow\rangle$. Now $\psi$ is a superposition of
$|\uparrow\rangle and |\downarrow\rangle$ which have opposite
accelerations. In particular if either A or B is zero, then the
expectation value of the acceleration operator vanishes; yet $\psi$ is
an equal superposition of these two eigenmodes with opposite
accelerations. This is analogous to the Stern-Gerlach experiment with
the spin perpendicular to the inhomogeneity of the magnetic field for
which the wave packet splits into the two states that are eigenstates
of the acceleration operator even though the classical acceleration
vanishes in this case$^9$.  If the dipole were charged, as in the case
of an electron, its
acceleration could be detected  by the radiation it emits.
For the present case of the neutron, it was shown how the
quantum mechanical magnetization and polarization caused by its
magnetic moment generates an electromagnetic field$^1$. Therefore
by measuring the electromagnetic
field associated with the neutron, its acceleration could in principle  be
detected,
even when it is in a plane wave state$^{10}$.

In a plane wave state, however, we cannot see the effect of the
acceleration as a motion of the wave because it extends uniformly
over all
space. Thus it is necessary to form a localized wave packet to observe the
acceleration in this manner. For a given ${\bf p}$, the
two eigenmodes with their spin states having quantization axis along
${\bf  B'}$ have vanishing expectation value for the acceleration,
because their spins are conserved. However, the plane wave components
forming the wave packet with different values of ${\bf p}$ have {\it
different directions} for the corresponding ${\bf  B'}$. Therefore the
spins for the different plane wave components with a fixed initial
polarization are not conserved, which is why the the wave packet is
able to accelerate  This is basically because of the velocity dependent
potential in (1).

To remove any remaining doubt, one can obtain the spatial motion
of a polarized neutron, whose initial
spatial wave function and spin polarization are arbitrary,
in constant electric and magnetic fields.
The wave function $\Psi$ at time $t = 0$ is taken to be
 $$\Psi({\bf x},0) =
\psi({\bf x})\phi$$
where $\phi$ is a normalized  two component constant spinor. Then
$$\Psi({\bf x},t) = exp(-itH)\Psi({\bf x},0)$$
where $H$ is given by (1). Then for small t, on using the Baker-
Campbell-Hausdorff formula,
$$\eqalign{\Psi({\bf x},t)=&exp(it{\mu\bf  \sigma}
\cdot {\bf  B})\,exp(-it {1 \over mc}{\mu{\bf  \sigma}} \cdot {\bf  p}
\times {\bf  E})\cr
& exp\{it^2{\mu^2\over mc}[({\bf  \sigma} \times{\bf
B})\times {\bf  E}]\cdot{\bf p}\}\psi_0({\bf x},t)\phi\quad ,}\eqno(13)$$
where the exponent of the last exponential arises from the
commutator of the last two terms in (1), with the higher order
commutators neglected, and
$$\psi_0({\bf x},t)=exp(-it{1 \over 2m} {\bf p}^2)\psi({\bf x})$$
is the wave packet at time t that would have evolved from
$\psi({\bf x})$ at $t=0$ in the absence of any external fields.

A postselection is now performed by
projecting this state on the spin state resulting from the Larmor
precession of $\phi$ during the time t [namely,
$\chi(t) = exp(it\mu{\bf \sigma}\cdot{\bf B})\phi]$,
by sending the beam through a Stern-Gerlach apparatus with a
magnetic field whose inhomogeneity is in the direction of
polarization of $\phi$. The probability amplitude for transition to
$\chi(t)$ is
$$\langle\chi(t)|\Psi({\bf x},t)\rangle=
\psi_0({\bf x}'(t),t)\quad ,\eqno(14)$$
for small t, where
 $${\bf x}'(t) = {\bf x} +t{\mu\over mc}
\langle{\bf \sigma}\rangle\times{\bf E} + t^2{\mu^2\over mc}
(\langle{\bf \sigma}\rangle\times{\bf B})\times{\bf E}$$
with $\langle{\bf \sigma}\rangle=\phi^\dagger{\bf \sigma}\phi$ and
the fact that $i{\bf p}$ in Eq. (13)
 is the translation operator has been used. Hence
the wave packet acquires the shift in velocity $-{\mu\over
mc}\langle{\bf \sigma}\rangle\times{\bf E}$, and acceleration
$-{2\mu^2\over mc}
(\langle\bf \sigma\rangle\times{\bf B})\times{\bf E}$ as predicted
by (2) and (5).
This can be tested in
principle by a polarized neutron beam striking a screen, turning on
the electromagnetic field, and observing the shift in the intensity
distribution on the screen. By repeating this experiment with
different electric and magnetic fields, it can be determined whether these
shifts are in agreement with the prediction (14). Thus one  sees
explicitly by means of a wave packet that the acceleration
predicted previously$^{1,2}$ is in principle observable for a spin 1/2
particle contrary to the conclusion of ref. 6.

J. A. thanks Y. Aharonov for clarifying discussions, and the NSF for
support under grant no. PHY-8807812. C. R. H. acknowledges the
support of U.S. Department of Energy Grant
No. DE-FG02-91ER40685.
\vskip 2cm
\noindent {\bf  \underbar{References}}
\item{1.} J. Anandan, Phys. Lett. A {\bf  138,} 347 (1989); errata
Phys. Lett. A. {\bf  152,} 504 (1991).
\item{2.} J. Anandan in Proc. 3rd Int. Symp. Found. of Quant. Mech.,
Tokyo, 1989, edited by S. Kobayashi {\it et al}. (Physical Society of Japan,
1990)
\item{3.} J. Schwinger, Phys. Rev. {\bf  73,} 407 (1948).
\item{4.} J. Anandan, Phys. Rev. Lett., {\bf  24,} 1660 (1982).
 \item{5.} W. Shockley and R. P. James, Phys. Rev. Lett. {\bf  18,}
876 (1967);
S. Coleman and J. H. Van Vleck, Phys. Rev. {\bf  171,} 1370 (1969);
Y. Aharonov and A. Casher, Phys. Rev. Lett. {\bf  53,} 319 (1984);
Y. Aharonov, P. Pearle, and L. Vaidman, Phys. Rev. A {\bf  37,} 4052
(1988).
\item{6.} R. C. Casella and S. A. Werner, Phys. Rev. Lett.
{\bf  69,} 1625 (1992).
\item{7.}A. Cimmino, G. I. Opat, A. G. Klein, H. Kaiser, S. A. Werner, M.
Arif, and R. Clothier, Phys. Rev. Lett. {\bf 63,} 380 (1989).
\item{8.}J. Anandan in {\it Topological Properties and Global Structure of
 Space-Time}, eds. P. G. Bergmann and V. De Sabbata (Plenum Press, NY 1985), p.
 1-14.
\item{9.}Another example, pointed out to us by Helmut Rauch, is the
interference of neutron beams around a line charge with the
neutrons polarized so that the {\it expectation value} of the force
vanishes, which was considered by Aharonov and Casher$^5$. As in
the Stern-Gerlach experiment described here, the beam splits as a
consequence of
the quantum mechanical force. This is unlike the Aharonov-Bohm
phase shift in which
there is a phase shift even though the acceleration (or force) operator
vanishes along the beams, and not merely its expectation value.
\item{10.}In support of this conclusion Y. Aharonov has provided us
with the following interesting argument. Consider two charged
particles that are EPR correlated in position and momentum and have
opposite accelerations. If a measurement is performed on one
particle so that it is localized, then the other would also be localized
and consequently each
particle would radiate. On the other hand if a measurement of
momentum is made on the first particle then each particle would be
in a plane wave state. Thus a plane wave state must also radiate since
 otherwise a signal could be sent faster than the speed of light.

\bye